\documentclass[journal=jctcce,type=communication]{achemso}
\setkeys{acs}{maxauthors=0,articletitle=true}

\usepackage{achemso}
\usepackage{graphics}
\usepackage{amssymb,amsfonts}
\usepackage{graphicx}
\usepackage[table]{xcolor}
\usepackage{multirow}
\usepackage{caption}
\usepackage{booktabs}
\usepackage{amsmath}
\usepackage{amsopn}
\usepackage{siunitx}
\usepackage{bm}
\usepackage{color}
\usepackage{ulem}

\DeclareMathOperator{\tr}{Tr}

\author{Janus J. Eriksen}
\email{janus.eriksen@bristol.ac.uk}
\affiliation{School of Chemistry, University of Bristol, Cantock's Close, Bristol BS8 1TS, United Kingdom}
\author{J{\"u}rgen Gauss}
\email{gauss@uni-mainz.de}
\affiliation{Department Chemie, Johannes Gutenberg-Universit{\"a}t Mainz, Duesbergweg 10-14, 55128 Mainz, Germany}

\title[TITLE]{Incremental Treatments of the Full Configuration Interaction Problem}

\begin{document}

%
%
\begin{abstract}

The recent many-body expanded full configuration interaction (MBE-FCI) method is reviewed by critically assessing its advantages and drawbacks in the context of contemporary near-exact electronic structure theory. Besides providing a succinct summary of the history of MBE-FCI to date within a generalized and unified theoretical setting, its finer algorithmic details are discussed alongside our optimized computational implementation of the theory. A selected few of the most recent applications of MBE-FCI are revisited, before we close by outlining its future research directions as well as its place among modern near-exact wave function-based methods.

\end{abstract}

\newpage

%
%

%
\section{Introduction}\label{intro_sect}

In a standard basis of Slater determinants, the exact solution to the time-independent electronic Schr{\"o}dinger equation in a finite one-electron basis will correspond to a linear, weighted sum over all possible electron configurations. This particular wave function Ansatz, known conventionally as full configuration interaction~\cite{knowles_handy_fci_cpl_1984,knowles_handy_fci_cpc_1989,knowles_fci_cpl_1989,knowles_handy_fci_jcp_1989,olsen_fci_jcp_1988,olsen_fci_cpl_1990} (FCI), has traditionally held a prominent and pivotal role in the field of quantum chemistry as the ultimate benchmark against which to compare and calibrate most of electronic structure theory. Crucial, though, is the fact that its formal exactness necessarily comes at the price of factorial computational complexity due to the combinatorial exercise associated with distributing $N$ electrons among $M$ molecular orbitals (MOs). In extended (or even non-minimal) basis sets, however, the overwhelming majority of the FCI wave function will be vastly redundant~\cite{ivanic_ruedenberg_ci_deadwood_tca_2001,bytautas_ruedenberg_ci_deadwood_cp_2009}. This empirical observation has thus served to drive a number of so-called selected CI (SCI) approximations over the years, which all seek to lift any of such configurational sparsity by focussing on energetically important determinants only~\cite{malrieu_cipsi_jcp_1973}. Depending on which particular measure of importance is being used, as well as whether the selection of determinants gets based on deterministic, stochastic, or even regressive protocols, the manifold of resulting methods will ultimately come to differ from one another in their subtleties and efficacies. For a comprehensive account of such modern takes on FCI, the reader is referred to a recent perspective on the matter by one of us~\cite{eriksen_fci_perspective_jpcl_2021}, which covers early, present, and near-future state-of-the-art approaches in greater detail than we will possibly be able to afford in the present review.\\

As a radically different approach to FCI, we will herein be concerned with its realizations by means of many-body expansions (MBEs). On that note, we will first and foremost discuss so-called MBE-FCI theory alongside its various incarnations. The initial work on modern MBE-FCI appeared in the year 2017~\cite{eriksen_mbe_fci_jpcl_2017} and its generalized methods have been in active development by us ever since~\cite{eriksen_mbe_fci_weak_corr_jctc_2018,eriksen_mbe_fci_strong_corr_jctc_2019,eriksen_mbe_fci_general_jpcl_2019,eriksen_mbe_fci_prop_jcp_2020}. Coincidentally, yet entirely unrelated, Zimmerman proposed a comparable theory around the same time, known as incremental FCI (iFCI)~\cite{zimmerman_ifci_jcp_2017_1}, which too continues to be extended to this date~\cite{zimmerman_ifci_jcp_2017_2,zimmerman_ifci_jpca_2017,zimmerman_ifci_jcp_2019}. As part of the present work, we will reiterate what common traits and discrepancies exist in-between these two theoretical frames. At an overall level, though, both MBE-FCI and iFCI constitute incremental approximations to FCI and thus fundamentally draw on Nesbet's original work on so-called Bethe-Goldstone theory from the 1960s~\cite{nesbet_phys_rev_1967_1,nesbet_phys_rev_1967_2,nesbet_phys_rev_1968}. However, generalized MBE-FCI may conveniently be viewed as the unifying umbrella-like setting, within which all of the discussed orbital-based incremental methods can be derived and expressed. The methods themselves are intended as pragmatic ways of circumventing the prohibitively exponential scaling wall of FCI by simulating properties without recourse to the exact, $N$-dimensional wave function~\bibnote{It is worth noting that Stoll have also touched upon the idea of using orbital-based MBEs for calculating correlation energies of solid-state systems, see, e.g., Refs. \citenum{stoll_cpl_1992,stoll_phys_rev_b_1992,stoll_jcp_1992}}. Conventionally, ground-state energies have been the desired property, but recent work has illustrated how MBEs may also facilitate the incremental calculation of other properties, not only for ground states, but rather for any state of arbitrary spin. Given this versatility of incremental approaches to FCI, it is our belief that methods based on orbital-based MBEs will continue to make a lasting impact in the electronic structure community, despite the fact that they are possibly more costly than their SCI counterparts. We will close our review by conveying a number of outstanding challenges in MBE-FCI as well as possible near-future solutions.

%
%

%
\section{Theory}\label{theory_sect}

Provided with a complete set of MOs, e.g., resulting from a preceding mean-field calculation---such as, a restricted Hartree-Fock (RHF) or a complete active space self-consistent field~\cite{roos_casscf_acp_1987,olsen_casscf_ijqc_2011} (CASSCF) treatment---an MBE-FCI calculation commences by deciding upon a division of these orbitals into two distinct and non-overlapping spaces, namely, a {\textit{reference}} and an {\textit{expansion}} space. The former of these, which, in turn, determines the latter in full, may be chosen to have any composition; if set to span the entirety of the space of virtual orbitals, MBE-FCI will coincide with the Nesbet's original theory and---subject to further truncation approximations of the reference space---also the iFCI method, while if the reference space holds all occupied orbitals of a given system, the method will be that of our initial work on virtual orbital MBEs~\cite{eriksen_mbe_fci_jpcl_2017}. However, for any choice of reference space different from these limiting cases, the basic theory is entirely generalizable to all conceivable correlation domains.\\

An MBE-FCI calculation now proceeds by performing an orbital-based MBE in the resulting expansion space, which will serve to recover the residual correlation missing from performing an initial complete active space configuration interaction (CASCI) calculation in the reference space alone. Denoting this (truncated FCI) reference correlation energy as $E_{\text{ref}}$, an MBE-FCI decomposition of the FCI correlation energy will formally read as
\begin{align}
E_{\text{FCI}} &= E_{\text{ref}} + \sum_{p}\epsilon_{p} + \sum_{p<q}\Delta\epsilon_{pq} + \sum_{p<q<r}\Delta\epsilon_{pqr} + \ldots \nonumber \\
&\equiv E_{\text{ref}} + E^{(1)} + E^{(2)} + E^{(3)} + \ldots + E^{(M_{\text{exp}})} \label{mbe_eq}
\end{align}
where the MOs of the expansion space (of size $M_{\text{exp}}$) of unspecified occupancy are labelled by indices $\{p,q,r,s,\ldots\}$, and $\epsilon_{p}$ designates the correlation energy of a CASCI calculation in the composite space of orbital $p$ and all of the MOs of the assigned reference space, with the reference correlation energy, $E_{\text{ref}}$, subtracted. At an arbitrary order $k$, the orbital increments, $\Delta\epsilon_{[\Omega]_{k}}$, are recursively defined for a general tuple of $k$ MOs, $[\Omega]_{k}$, via the following relation
\begin{align}
\Delta\epsilon_{[\Omega]_{k}} = \epsilon_{[\Omega]_{k}} - \big(\sum_{p \in S_1[\Omega]_{k}}\epsilon_{p} + \sum_{pq \in S_2[\Omega]_{k}}\Delta\epsilon_{pq} + \ldots + \sum_{pqrs\cdots \in S_{k-1}[\Omega]_{k}}\Delta\epsilon_{pqrs\cdots}\big) \ . \label{increment_eq}
\end{align}
In Eq. \ref{increment_eq}, the action of the operator $S_{l}$ onto $[\Omega]_{k}$ is to yield all possible unique subtuples of order $l$ ($1 \leq l < k$), and $\epsilon_{[\Omega]_{k}}$ is defined on par with $\epsilon_{p}$ above in Eq. \ref{mbe_eq}.\\

The most general treatment of electron correlation is now achieved by choosing upon an empty reference space, in which case the total expansion space will span all of the MOs of the system at hand. As for when the reference space encompasses all virtual or occupied MOs, $E_{\text{ref}}$ vanishes in this case, but the expansion in Eq. \ref{mbe_eq} will now commence at second rather than first order with all possible unique pairs of correlating occupied and virtual orbitals~\bibnote{The reason for the absence of first-order contributions in the case of an empty reference space is simply that a CASCI calculation in the space of either a single occupied or virtual orbital will miss a corresponding correlation space to excite electrons into or miss any electrons to excite, respectively.}. However, despite the fact that MBE-FCI will be free of any constraints in this case, it will importantly not represent a true {\textit{tabula rasa}} approach to FCI as an implicit bias still exists in the choice of MO basis. To that end, Stoll has recently presented work on the optimization of localized orbitals at a partially correlated level, rather than for the traditional uncorrelated HF starting point~\cite{stoll_jcp_2019}. In our experience hitherto, spatially localized orbitals often represent an excellent choice of MO basis within the scope of MBE-FCI as they allow for more compressed expansions (fewer significant contributions altogether), in turn, leading to faster convergence profiles. Another typical choice is to use a set of orbitals tailored to a specific correlated method~\cite{lowdin_nat_orb_fci_phys_rev_1955,lowdin_nat_orb_fci_phys_rev_1956}. We have previously proposed the use of these MOs in combination with a so-called base model in the expansion~\cite{eriksen_mbe_fci_weak_corr_jctc_2018}. In this case, rather than the total correlation energy, one expands the gap between FCI and an intermediate correlated model, for which a preliminary calculation on the whole system is feasible, in part, also to compute the one-electron reduced density matrix needed to derive its natural orbitals (NOs). As the quantity that need be recovered by the MBE will be significantly reduced, the use of base models can lead to faster convergences. However, additional constraints are tied into the MBE in this case as the base model is fundamentally required to yield a reasonable approximation to FCI for it to accelerate the underlying MBE-FCI algorithm. In cases where this is not true, e.g., in the presence of static and strong correlation, typical choices of base models, e.g., the acclaimed CCSD(T) method of coupled cluster theory~\cite{original_ccsdpt_paper}, will generally be limited in their performance as they are themselves based on a single determinant.\\

Now, in the case where a given system is indeed dominated by a single reference determinant (say, the HF determinant), this determinant will have the largest weight in the linear FCI expansion of the wave function and the system is said to be dominated by dynamical correlation alone. However, in cases where more than one determinant have significant weights, one will need to be able to describe the important (and common) constituents of these from the beginning and they will then need be included in the reference space. More generally, capturing the integral part of the total electron correlation in the reference space will lead to a faster convergence of the expansion. This observation will hold true even when the HF determinant dominates the FCI wave function, and reducing the size of the expansion space will additionally lead to fewer possible orbital tuples throughout the MBE. For linear molecules, the $\pi$-pruning technique of Refs. \citenum{eriksen_mbe_fci_strong_corr_jctc_2019} and \citenum{eriksen_mbe_fci_prop_jcp_2020} has further been introduced in order to deal with the $D_{\infty\text{h}}$ and $C_{\infty\text{v}}$ point groups within their $D_{2\text{h}}$ and $C_{2\text{v}}$ subgroups, respectively. Essentially, $\pi$-pruning is a sort of prescreening filter that works to prune away all increment calculations that fail to simultaneously include the $x$- and $y$-components of a given pair of degenerate $\pi$-orbitals. The use of $\pi$-pruning generally results in much shorter (faster) expansions for molecules belonging to non-Abelian point groups, while at the same time warranting convergence onto states spanned by the correct irreducible representation. While $\pi$-pruning is a specific filter designed for the treatment of a specific type of systems, it also serves as an example of how it is generally possible to add alternative filters in a top-down manner in order to accelerate or assure convergence onto a target FCI property.

%
%

%
\section{Implementation}\label{implementation_sect}

As a platform for our theoretical work on MBE-FCI over the past couple of years, we have developed our own Python-based, open-source {\texttt{PyMBE}} code~\cite{pymbe}. All electronic structure kernels within {\texttt{PyMBE}} are formulated upon the {\texttt{PySCF}} program~\cite{pyscf_wires_2018,pyscf_jcp_2020}, and the {\texttt{MPI4Python}} module handles parallel communication over the message passing interface (MPI) standard~\cite{mpi4py_1,mpi4py_2,mpi4py_3}.\\

Ever since its conception, the {\texttt{PyMBE}} code has seen several rounds of heavy optimizations. In particular, the memory handling and footprint of the involved 1- and 2-electron integrals as well as all involved intermediates and results have been overhauled. The recursive nature of Eq. \ref{increment_eq} implies that we need to be able to look up a vast number of subset contributions when calculating a given increment. In {\texttt{PyMBE}}, this is achieved by representing every tuple of orbitals by its hash and using binary searches in finding all subset occurrences. However, this setup implies that a sorted array of hashes must be stored in addition to the actual increments which they correspond to, adding to the memory requirements in the code. For that reason, a hybrid MPI+MPI approach to shared-memory allocation and access has been implemented throughout our code, in which the {\texttt{MPI$\_$Win$\_$allocate$\_$shared}} function is used as a departure from the standard abstract and distributed memory model of MPI~\cite{hoefler_mpi_mpi_computing_2013}. As previously detailed in Ref. \citenum{eriksen_mbe_fci_general_jpcl_2019}, the underlying memory organisation on a given computer node gets exposed to MPI, allowing us, in turn, to bypass the expensive and convoluted MPI-3 one-sided operations by instead using shared memory directly between the processes on the node. The effect of this model is that MPI is employed in markedly different manners across and within individual nodes. For an MBE-FCI calculation on $A$ number of nodes---each equipped with $B$ cores---our parallel model comprises a single global master and $AB-1$ global slaves. Among these, $A-1$ slaves are additionally promoted to local masters that pass messages onto the global master via traditional distributed MPI, while each sharing a window to the physical memory address space on their respective node with their own group of $B-1$ local slaves. The latter communication proceeds over dedicated communicators.\\

At any given order in the MBE, all possible orbital tuples are being yielded by a generator function, which takes into account the composition of the reference space. In case this is empty, only tuples that make reference to both occupied and virtual orbitals are allowed, as touched upon in Section \ref{theory_sect}. The task scheduling has been implemented in a round-robin fashion and the input generator has been designed in such a way that those tuples that correspond to large (determinant-wise) CASCI calculations will precede the smaller ones. Once a given process has been assigned a tuple (i.e., a unique work task), it proceeds by computing the specific core and active space indices needed for the CASCI calculation. Next, the corresponding 1- and 2-electron integrals are extracted alongside the core energy. In {\texttt{PyMBE}}, these are stored in the transformed MO basis in shared memory with the electron repulsion integrals compressed into a matrix form with 4-fold symmetry. Finally, the CASCI energy is calculated, before the increment is computed and stored alongside the corresponding hash. If enabled, a full suite of restart functionalities guarantee that MBE-FCI is trivially protected against hardware failures, strict time limits, etc., at only a minimum of associated penalty.\\

The most recent version of our screening protocol has been implemented such that MOs get gradually screened away from the full expansion space according to the absolute magnitude of the tuples which they take part in. Specifically, a certain percentage of the MOs that contribute the least may get removed (governed by a dedicated input parameter). As shrinking the expansion space is accompanied by a corresponding reduction in the number of increment calculations at the orders to follow, only those specific MOs of the expansion space (at any given order) that give rise to the numerically largest increments will be retained among the tuples at the following order. In the case one or more orbitals get screened away, the code thus enters a recently implemented purging module, which is designed to retain only those contributions at lower orders that are needed going forward. The rationale behind this step in an MBE-FCI calculation is that all subsequent tuples will not make reference to the screened orbitals and their increments at lower orders are hence not required anymore. This purging procedure generally works to lower our memory requirements significantly.\\

\begin{figure}[ht]
\begin{center}
\includegraphics[width=0.85\textwidth]{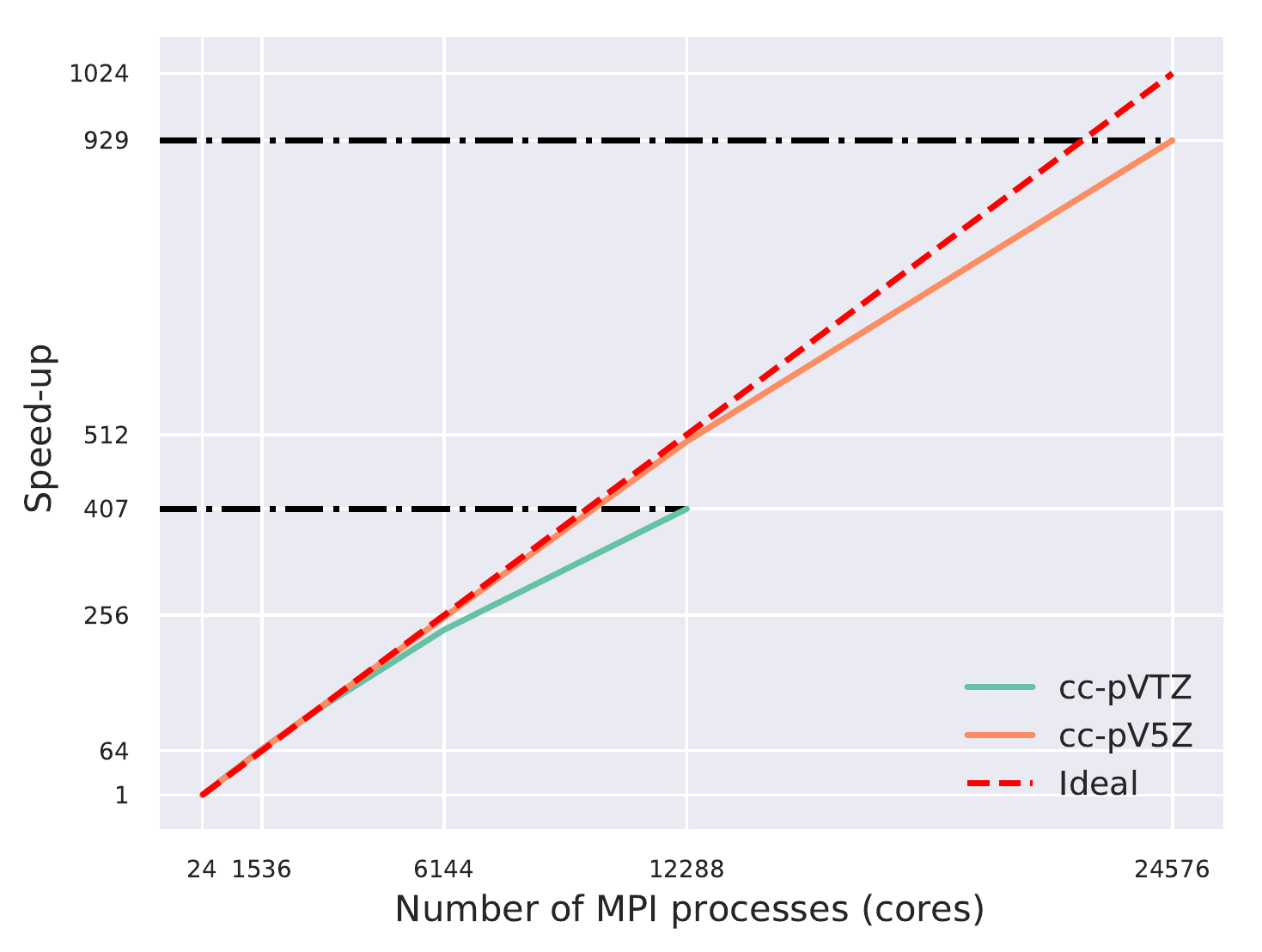}
\caption{Internode strong scaling of frozen-core MBE-FCI calculations on H$_2$O~\cite{eriksen_mbe_fci_strong_corr_jctc_2019}.}
\label{h2o_parallel_scaling_fig}
\end{center}
\vspace{-0.6cm}
\end{figure}
The parallel scaling potential of MBE-FCI was assessed in Ref. \citenum{eriksen_mbe_fci_strong_corr_jctc_2019}. Being compute- rather than memory-bound, the resource utilization of the theory and its implementation within the {\texttt{PyMBE}} code is best measured in terms of its strong scalability. In Figure \ref{h2o_parallel_scaling_fig}, the relative speed-up gained by moving from a single node (Intel Xeon E$5$--$2680$ $\text{v}3$ hardware with $24$ cores {@} $2.5$ GHz and $128$ GB of global memory), on which MPI is employed across all of its 24 cores, to a total of 512/1,024 nodes (i.e., 12,288/24,576 individual MPI processes) is shown. The scalability in Figure \ref{h2o_parallel_scaling_fig} has been assessed for MBE-FCI calculations on the standard H$_2$O molecule in medium-sized (cc-pVTZ) and extended (cc-pV5Z) basis sets~\cite{dunning_1_orig}. At scale, the efficiencies at 512 (12,288 cores) and 1,024 nodes (24,576 cores) amount to $79\%$ and $91\%$ for the expansions in the cc-pVTZ and cc-pV5Z basis sets, respectively, and the difference in performance between the two basis sets can be ascribed to the significantly larger number of individual CASCI calculations in the latter of the two expansions. MBE-FCI is thus seen to offer a highly scalable treatment of the electron correlation problem with a massive parallelism that is ideally suitable for modern distributed supercomputers.

%
%

%
\section{Applications}\label{applications_sect}
\begin{figure}[ht]
\begin{center}
\includegraphics[width=0.85\textwidth]{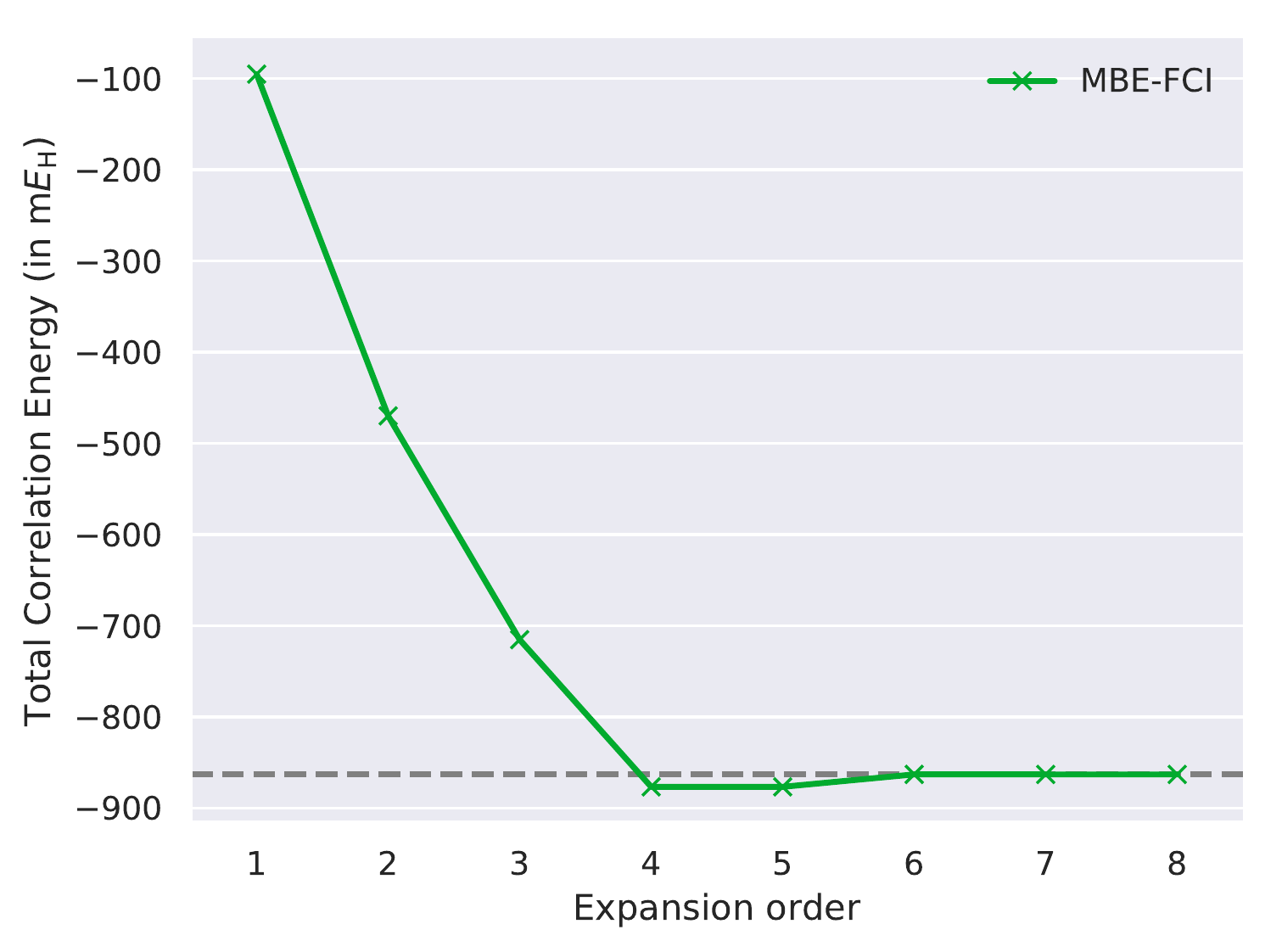}
\caption{MBE-FCI convergence for the frozen-core C$_6$H$_6$/cc-pVDZ problem with localized MOs and an ($6e$,$6o$) reference space~\cite{eriksen_benzene_jpcl_2020}. The dashed line indicates $\Delta E = -863$ m$E_{\text{H}}$. (Reproduced with permission from Ref. \citenum{eriksen_benzene_jpcl_2020}. Copyright 2020, American Chemical Society)}
\label{c6h6_fig}
\end{center}
\vspace{-0.6cm}
\end{figure}
Having covered its theoretical basis, we will now review a selected few of the molecular systems for which MBE-FCI has been applied to date. Most recently, MBE-FCI took part in a broad, international blind challenge devoted to computing the best possible estimate of the FCI correlation energy for the ubiquitous benzene system in a standard correlation consistent double-$\zeta$ basis set~\cite{eriksen_benzene_jpcl_2020}. The repertoire of high-accuracy methods that entered the challenge was collectively chosen to constitute a truly diverse view of leading, contemporary approaches from all around the world, and, as such, the work was not just testament to what is currently achievable by means of near-exact quantum chemistry, but also to what the future beholds. In the specific case of benzene, and with respect to how to optimally apply MBE-FCI to the problem of estimating its correlation energy, it is imperative to initially note that one does not necessarily need to include any orbitals in the reference space at all, as was previously demonstrated in Ref. \citenum{eriksen_mbe_fci_general_jpcl_2019}. However, doing so may potentially lead to faster convergence. For this reason, the MBE-FCI entry in Ref. \citenum{eriksen_benzene_jpcl_2020} was computed using a reference space that included the six frontier bonding and anti-bonding valence $\pi$-orbitals in a localized Pipek-Mezey (PM) basis~\cite{pipek_mezey_jcp_1989}. The convergence of the final result is presented in Fig. \ref{c6h6_fig} ($\Delta E = -863.03$ m$E_{\text{H}}$), in essentially perfect agreement with what is nowadays regarded as the best estimate of the correlation energy on the basis of the blind challenge.\\

The MBE-FCI calculation shown in Fig. \ref{c6h6_fig} involved in excess of 1.55 billion individual CASCI calculations of variable size and composition and consumed a staggering 1.5 million core-hours in the process (on Intel Xeon E5-2697v4 (Broadwell) hardware with 36 cores $@$ 2.3 GHz and 3.56 GB/core of physical memory, Galileo supercomputer at CINECA (Italy)). Although improvements have subsequently been made to the code base---to the extent where the total compute time can be reduced by close to a factor 2 (cf. Section \ref{implementation_sect})---the theory behind MBE-FCI is arguably somewhat more expensive than its alternatives. However, with these exhaustive resource requirements follows rigour, as evidenced by how well converged the final energy is. The change in energy across the final two orders in the expansion amounts to a mere $-0.04$ m$E_{\text{H}}$, or $-0.1$ kJ/mol, that is, well within thermochemical tolerance. However, and this is important to emphasize, no methodical measure of the final uncertainty against FCI currently exists, which we will elaborate further on in Section \ref{outlook_sect}. That being said, given ample computational resources, even larger systems (of a similar nature) will also be amenable to a treatment by MBE-FCI as the dimension of the largest CASCI calculations in Fig. \ref{c6h6_fig} is well within the capabilities of even today's optimized FCI kernels.\\

\begin{figure}[ht]
\begin{center}
\includegraphics[width=0.85\textwidth]{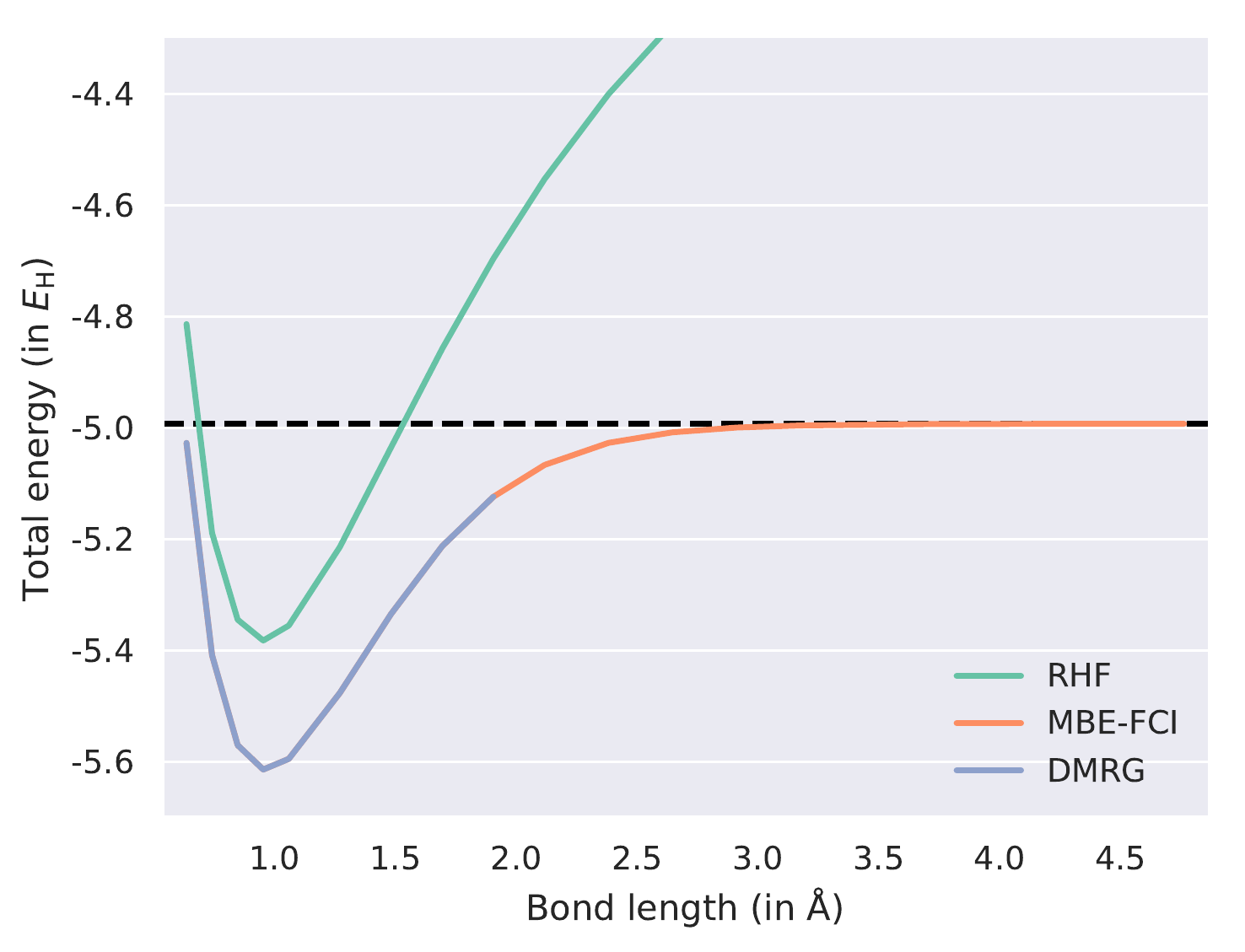}
\caption{RHF, MBE-FCI, and DMRG results for the dissociation curve of an H$_{10}$ chain in a cc-pVDZ basis set~\cite{eriksen_mbe_fci_strong_corr_jctc_2019,simons_collab_bond_break_hydrogen_prx_2017}. The dashed line indicates the FCI complete dissociation limit.}
\label{h10_chain_dz_pm_fig}
\end{center}
\vspace{-0.6cm}
\end{figure}
As an example of how to deal with static correlation, MBE-FCI was applied in Ref. \citenum{eriksen_mbe_fci_strong_corr_jctc_2019} to the popular problem of simultaneous stretching all bonds of a tenfold linear chain of hydrogen atoms. For this model system, the usual transition from weak to strong correlation observed along practically all bond dissociations takes place, albeit on a significantly more extended scale as the involved bonds are all broken at once. While MBE-FCI based on a CASSCF(10,10) reference was shown to successfully reproduce reference results obtained using density matrix renormalization group (DMRG) theory~\cite{simons_collab_bond_break_hydrogen_prx_2017}, MBE-FCI will hold limited promise of providing detailed information in the thermodynamic limit whenever it has to rely on CASSCF($N$,$N$) expansion references. To that end, as was discussed in detail in Ref. \citenum{eriksen_mbe_fci_strong_corr_jctc_2019}, canonical CASSCF orbitals are bound to remain delocalized over large sections of the chain, a fact which in turn inhibits the orbital screening. For this reason, additional results, using a reference space comprising only the RHF determinant, but localized PM rather than canonical virtual orbitals, were furthermore presented. Displayed in Fig. \ref{h10_chain_dz_pm_fig}, these results are in excellent agreement with DMRG for all but the shortest bond distances in the repulsive region where the concept of locality is anyways somewhat ill-defined. Importantly, due to the formulation on a standard RHF rather than a CASSCF reference, MBE-FCI is potentially transferrable to larger chains and basis sets (and even other topologies, e.g., rings and sheets), thus offering a viable approach for the treatment of the thermodynamic limit.\\

While its original formulation was focussed solely on the calculation of correlation energies for closed- and open-shell systems, MBE-FCI has recently been extended to the treatment of excitation energies and dipole moments for ground and excited states~\cite{eriksen_mbe_fci_prop_jcp_2020}. In analogy with Eq. \ref{mbe_eq}, excitation energies may be computed by an expansion of the energetic gap between the ground and an excited state, $E^{0n}$, rather than the correlation energy
\begin{align}
E^{0n}_{\text{FCI}} = E^{0n}_{\text{ref}} + \sum_{p}\epsilon^{0n}_{p} + \sum_{p<q}\Delta\epsilon^{0n}_{pq} + \sum_{p<q<r}\Delta\epsilon^{0n}_{pqr} + \ldots \label{mbe_ex_energy_eq}
\end{align}
As a CASCI calculation in an active space absent of any form of electron correlation will yield no correlation energy (comprising only the HF solution) and hence no excited states, $\epsilon^{0n}_{p}$ in Eq. \ref{mbe_ex_energy_eq} will be defined on par with $\epsilon^{0}_{p}$ in Eq. \ref{mbe_eq}. With respect to MBE-FCI for static properties, the calculation of these may be exemplified by focussing on electronic dipole and transition dipole moments. From the wave function coefficients of an individual CASCI calculation, the corresponding 1-electron reduced density matrix (RDM), $\bm{\gamma}^{n}$, for state $n$ may be readily computed, from which an electronic dipole moment is given as
\begin{align}
\bm{\mu}^{n}_{r} = -\tr[\bar{\bm{\mu}}_{r}\bm{\gamma}^{n}] \label{dipmom_eq}
\end{align}
in terms of dipole integrals, $\bar{\bm{\mu}}_{r}$, in the MO basis for each of the three Cartesian components, $r$. Letting these quantities take up the role of correlation or excitation energies in Eqs. \ref{mbe_eq} and \ref{mbe_ex_energy_eq}, respectively, results in the following decomposition of the FCI electronic dipole moment
\begin{align}
\bm{\mu}^{n}_{\text{FCI}} = \bm{\mu}^{n}_{\text{ref}} + \sum_{p}\bm{\mu}^{n}_{p} + \sum_{p<q}\Delta\bm{\mu}^{n}_{pq} + \sum_{p<q<r}\Delta\bm{\mu}^{n}_{pqr} + \ldots \label{mbe_dipole_eq}
\end{align}
Adding the nuclear component, $\bm{\mu}_{\text{nuc}} = \sum_{K}Z_{K}\bm{r}_{K}$, returns the molecular dipole moment. Finally, transition dipole moments, $\bm{t}^{0n}$, may be evaluated on par with Eq. \ref{mbe_dipole_eq}, except for the fact that the individual increments are computed on the basis of transition RDMs, $\bm{\bm{\gamma}}^{0n}$, which may be arrived at using the wave functions of both states involved in a given CASCI calculation. Being a vector rather than a scalar quantity, the screening procedure proceeds along all three Cartesian components ($x,y,z$) in the case of (transition) dipole moments and must be simultaneously fulfilled for all if a given MO is to be screened away from the expansion space.\\

\begin{figure}[ht]
\begin{center}
\includegraphics[width=.85\textwidth]{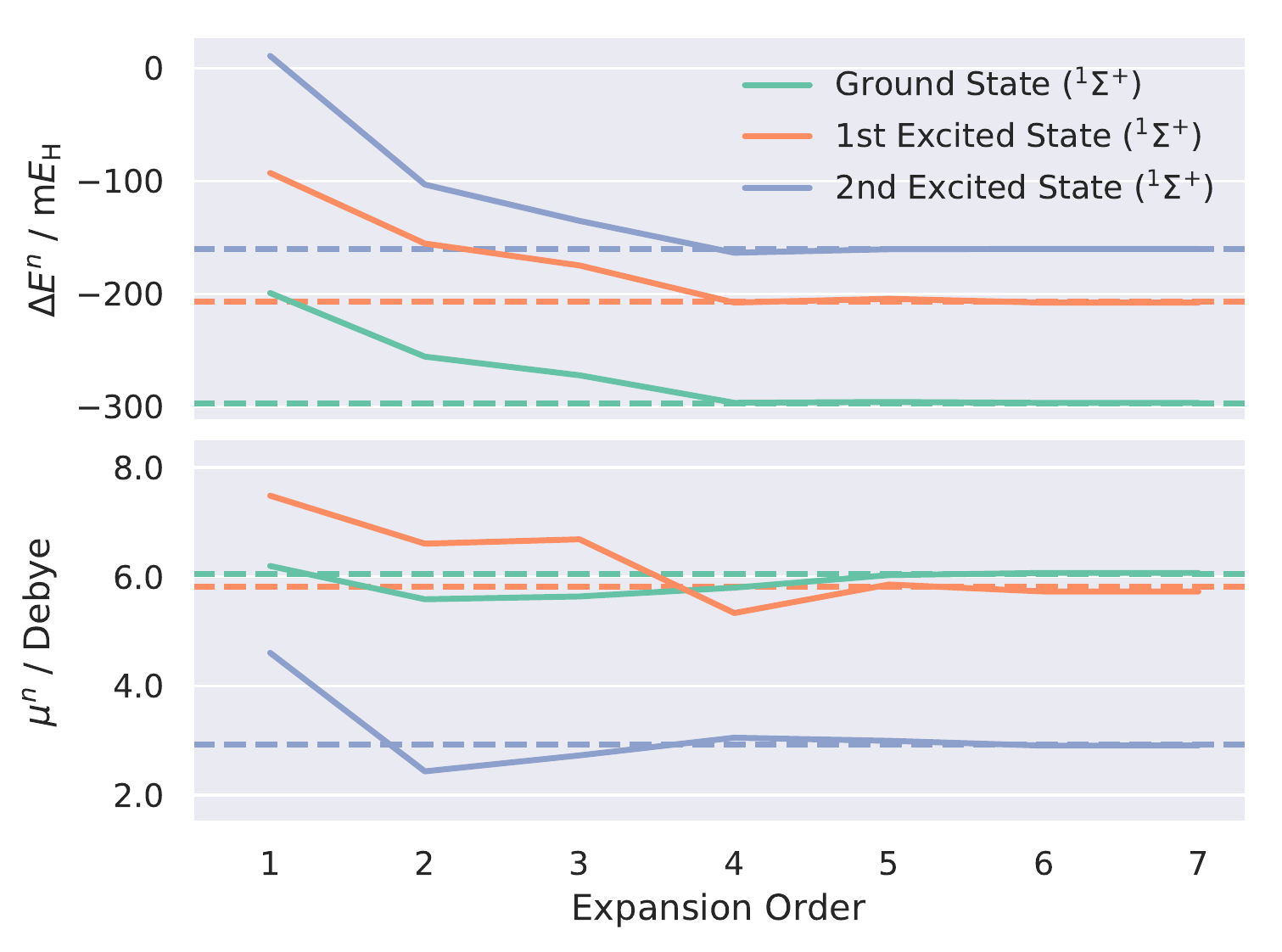}
\caption{Correlation energies ($\Delta E^{n}$, upper panel) and dipole moments ($\mu^{n}$, lower panel) for the ground and first two excited states (${^{1}}\Sigma^{+}$ symmetry) of MgO in the aug-cc-pVDZ basis set. Solid and dashed lines denote MBE-FCI and $i$-FCIQMC results, respectively~\cite{eriksen_mbe_fci_prop_jcp_2020,booth_alavi_fciqmc_jcp_2017}. (Reproduced with permission from Ref. \citenum{eriksen_mbe_fci_prop_jcp_2020}. Copyright 2020, AIP Publishing)}
\label{mgo_fig}
\end{center}
\vspace{-0.6cm}
\end{figure}
In Figure \ref{mgo_fig}, results are shown for the energies and dipole moments of the three lowest states of MgO in an aug-cc-pVDZ basis set, for which the frozen-core FCI correlation problem is described by the distribution of 16 electrons in 48 orbitals. All of the MBE-FCI results have been obtained using a reference space spanned by (state-averaged) CASSCF(8,8) calculations, and comparisons have been made with state-of-the-art FCI quantum Monte Carlo ($i$-FCIQMC) results from Ref. \citenum{booth_alavi_fciqmc_jcp_2017}. From the MBE-FCI results in Figure \ref{mgo_fig}, it is observed how the method correctly converges onto the individual states of interest. Despite the fact that the three states lie in close proximity of each other, MBE-FCI succeeds in distinguishing between them. In general, the convergence profiles in the individual plots of Fig. \ref{mgo_fig} are all different, as are the MO manifolds being screened away in the calculations, attesting to the fact that the different properties in question are inherently unrelated and that MBE-FCI is flexible enough to cope with this within an orbital-based expansion framework.\\

\begin{figure}[ht]
\begin{center}
\includegraphics[bb=7 7 424 312, width=0.85\textwidth]{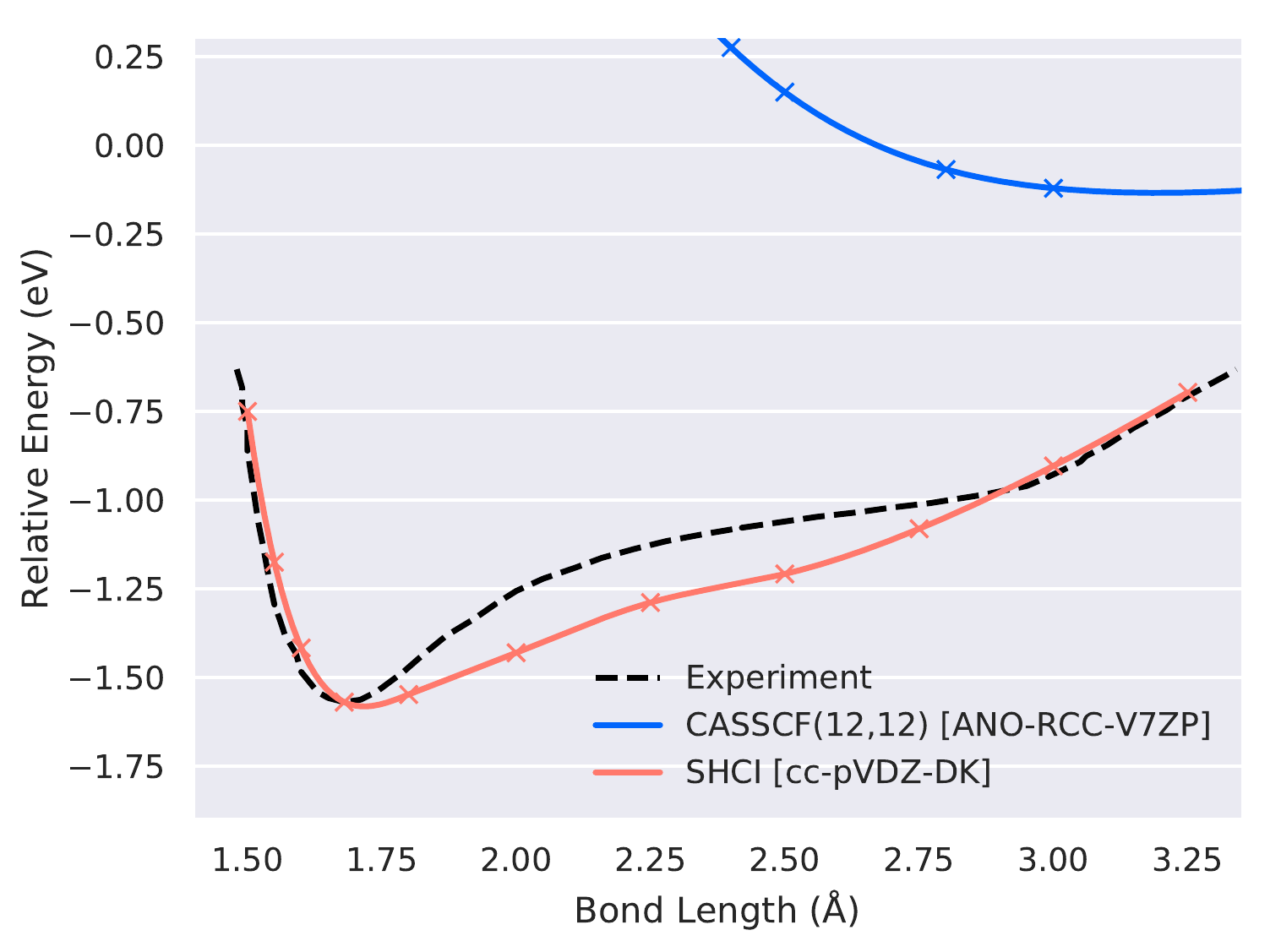}
\caption{Experimental and simulated results for the Cr$_2$ dissociation curve~\cite{casey_leopold_bond_break_cr2_1993,gagliardi_cr2_jctc_2016,sharma_umrigar_heat_bath_ci_prr_2020}.}
\label{cr2_pes_fig}
\end{center}
\vspace{-0.6cm}
\end{figure}
What are then the practical limitations of current-generation incremental approaches like MBE-FCI? Consider the potential energy curve of the chromium dimer in Fig. \ref{cr2_pes_fig}, which has nowadays developed into an appraised stress test for high-accuracy electronic structure theory. As per valence-bond theory, Cr$_2$ has a formal hextuple bond and the ${^{1}}\Sigma^{+}_{g}$ ground state will eventually dissociate into two equivalent atoms, each in a configuration of high spin with a total of 6 unpaired electrons in the Cr $3d$ and $4s$ atomic orbitals (AOs). Unlike for the H$_{10}$ chain in Fig. \ref{h10_chain_dz_pm_fig}, where the occupied (and to some extent even the virtual) MOs localize optimally onto the involved atomic centers, this is not necessarily the case in general, more complex system built from atoms of arbitrary covalency. For instance, in the present case of Cr$_2$, the 12 electrons in question, alongside the 12 MOs that map to the corresponding AOs, will demand special consideration along the bond dissociation coordinate. In the language of MBE-FCI, the smallest possible reference space, which would remain unaltered as the bonds are elongated, will thus coincide with this ($12$,$12$) active space. As CASSCF in this valence space only yields a very shallow minimum (at an unreasonable large bond length) for this particular system, the actual MBE in the expansion space, which is accountable for the general treatment of dynamical correlation, will soon come to involve excessively large CASCI calculations. In the vicinity of the equilibrium bond length, however, MBE-FCI has previously in Ref. \citenum{eriksen_mbe_fci_general_jpcl_2019} been shown to perform on par with the best possible reference methods~\cite{chan_dmrg_review_jcp_2015}, even when based on an empty reference space. However, upon elongating the bonds, the individual CASCI calculations of such an unbiased MBE start to become increasingly ill-conditioned and eventually fail to properly converge, as they cannot possibly accommodate the important (open-shell) determinants of the aforementioned minimal active space. For the specific case of the Cr$_2$ dissociation, alternative approaches like DMRG or even state-of-the-art semistochastic heat-bath CI (SHCI) are needed for a qualitatively correct description of the electronic structure in all correlation domains~\cite{sharma_umrigar_heat_bath_ci_prr_2020}. The SHCI results from Ref. \citenum{sharma_umrigar_heat_bath_ci_prr_2020} are reproduced in Fig. \ref{cr2_pes_fig}, but it is important to note that even this method is having its capabilities stretched in the shoulder region from $1.8$ to $2.7$ \AA, despite the use of a modest basis set of double-$\zeta$ quality and a scalar-relativistic Hamiltonian~\bibnote{Private correspondence with Cyrus Umrigar (Cornell University).}. Further to that, the basis set dependence exhibited by, e.g., CCSD(T) or SHCI is strong at all considered bond lengths, which explains the pronounced differences with respect to experiment data that are still visible in Fig. \ref{cr2_pes_fig}. As impressive as the SHCI results are on their own, the prospects of rationalizing the odd profile of the Cr$_2$ dissociation curve by means of near-exact quantum chemistry thus remain somewhat elusive for now.

%
%

%
\section{Outlook}\label{outlook_sect}

In comparison with most of the alternative methods in existence today, many of these are bound to allow for faster and perhaps more affordable routes towards simulating FCI properties than MBE-FCI, cf. Ref. \citenum{eriksen_fci_perspective_jpcl_2021}. However, MBE-FCI arguably sets itself apart from the rest by offering an incremental, robust, and widely applicable approach which is principally not restricted by the exponential scaling wall encountered in, e.g., the various SCI approaches. In addition, the flexibility of orbital-based MBEs admits a number of further knobs to turn over traditional approaches centred around individual determinants or configuration state functions; namely, besides variances with respect to the employed orbital basis, generalized MBE-FCI allows for the use of different reference spaces. In the asymptotic limit of an untruncated expansion, the choice of reference space will be irrelevant as the expansion trivially yields the exact FCI results, but upon introducing effective protocols for screening (incrementally) negligible orbitals away from the corresponding expansion spaces, some choices of reference spaces will yield noticeable more compact expansions than others. However, how to decide upon an optimal choice in as black-box a manner as possible still remains mostly unsolved. One feasible option may be to leverage information on independent orbital correlations at low orders in the MBE, but additional work on more automated selection schemes will be the topic of future work on extensions and refinements of MBE-FCI.\\

Moving forward, it will furthermore be interesting to see if regression techniques or related statistical processes may be implemented to learn certain components of MBE-FCI in lieu of a brute-force account of the underlying electronic structure all the way up throughout the MBE. For instance, despite the fact that efficient protocols have been implemented to screen away incremental contributions to the MBE that are deemed energetically redundant, these remain rather {\textit{ad hoc}} in the sense that they lack rigour and rely on simplified estimates of the correlation between MOs. A promising idea is now to use modern machine learning~\cite{dral_ml_qc_review_jpcl_2020,lilienfeld_burke_nat_commun_2020}, particularly models capable of disentangling the correlation patterns present among the individual MOs on the basis of the corresponding increments~\bibnote{Related ideas have previously also been proposed by Jason D. Goodpaster (University of Minnesota) in a lecture at the workshop {\textit{New Frontiers in Electron Correlation}}, Telluride, CO, US, in June 2019}. Given machine models for different systems, these will collectively aid in the design of transferable descriptors for use in all future MBE-FCI calculations. Not only will these enhancements of MBE-FCI see its runtime execution accelerated, but refined models may further be used to correct final results by terms that account for the most important of the screened increments. In turn, this will enable proxies for assessing the inherent uncertainty of an MBE-FCI run, something which is currently missing. While the use of regression for this purpose will necessarily result in a departure from the otherwise rigorous grounds of MBE-FCI, the importance of available error estimates cannot be underestimated, a point which is equally true in the case of other near-exact methods. This pertinent issue was also recently discussed in Ref. \citenum{eriksen_benzene_jpcl_2020}.\\

In extending MBE-FCI further, it is worth noting that the underlying theory is by no means restricted to FCI targets; for instance, orbital-based expansions may easily well be employed within coupled cluster theory. In addition, the use of alternative and approximative FCI solvers as a means to allow for larger (and faster) CASCI calculations remains to be explored within MBE-FCI. Finally, in the spirit of recent work seeking to revitalize the idea of transcorrelation~\cite{boys_handy_transcorr_prsoc_1969,handy_transcorr_jcp_1973,alavi_transcorr_jctc_2018,alavi_transcorr_prb_2019,alavi_transcorr_jcp_2019,reiher_transcorr_jcp_2020}, a natural, albeit non-trivial extension of the current generation of MBE-FCI will be to allow for expansions to spawn from a similarity-transformed Hamiltonian, akin to what is found in equation-of-motion coupled cluster theory~\cite{krylov_eom_cc_review_arpc_2008}. Its appealing traits as a correlated zeroth-order formulation of electron correlation aside, the non-Hermiticity of the theory will pose entirely new conceptual as well as technical challenges. Incorporating correlation directly into the very Ansatz of MBE-FCI, however, is bound to result in even faster convergent MBEs and may further prompt the development of a wealth of alternative methods. These may not necessarily be aimed at a rigorously defined target property (e.g., FCI ground- or excited-state energies or dipole moments), but rather seeking to profit from the fact that even low-order truncations of MBE-FCI will yield qualitatively accurate properties, as an alternative to contemporary correlated wave function theory.

%
%
\section*{Acknowledgments}

The authors wish to thank Cyrus Umrigar of Cornell University for fruitful discussions and for sharing part of the data behind Fig. \ref{cr2_pes_fig}.

%
%
\section*{Funding Information}

J.J.E. is grateful to the Alexander von Humboldt Foundation as well as The Independent Research Fund Denmark for financial support throughout the past couple of years.

%
%
\section*{Research Resources}

The authors acknowledge access awarded to the Galileo supercomputer at CINECA (Italy) through the $18^{\text{th}}$ PRACE Project Access Call as well as the Johannes Gutenberg-Universit{\"a}t Mainz for computing time granted on the MogonII supercomputer.

%
%
\section*{Data Availability Statement}

The data that support the findings of this study are available from the corresponding author upon reasonable request.

\providecommand{\latin}[1]{#1}
\makeatletter
\providecommand{\doi}
  {\begingroup\let\do\@makeother\dospecials
  \catcode`\{=1 \catcode`\}=2 \doi@aux}
\providecommand{\doi@aux}[1]{\endgroup\texttt{#1}}
\makeatother
\providecommand*\mcitethebibliography{\thebibliography}
\csname @ifundefined\endcsname{endmcitethebibliography}
  {\let\endmcitethebibliography\endthebibliography}{}


\begin{mcitethebibliography}{60}
\bibitem[Not()]{Note-2}
The reason for the absence of first-order contributions in the case of an empty reference space is simply that a CASCI calculation in the space of either a single occupied or virtual orbital will miss a corresponding correlation space to excite electrons into or miss any electrons to excite, respectively\relax
\mciteBstWouldAddEndPuncttrue
\mciteSetBstMidEndSepPunct{\mcitedefaultmidpunct}
{\mcitedefaultendpunct}{\mcitedefaultseppunct}\relax
\EndOfBibitem
\end{mcitethebibliography}
\end{document}